\documentclass[aps,prl,twocolumn]{revtex4}

\usepackage{graphicx}
\usepackage{amsmath}

\newcommand{\beq}{\begin{equation}}
\newcommand{\enq}{\end{equation}}
\newcommand{\bea}{\begin{eqnarray}}
\newcommand{\ena}{\end{eqnarray}}
\newcommand{\kbar}{\mathchar'26\mkern-9muk}
\newcommand{\heff}{\kbar}
\newcommand{\tav}{T_{\rm av}}

\begin{document}
\title{Ratchet effect for cold atoms in an optical lattice}
\author{Emil Lundh}
\affiliation{Department of Physics, Royal Institute of Technology,
AlbaNova, SE-106 91 Stockholm, Sweden}
\author{Mats Wallin}
\affiliation{Department of Physics, Royal Institute of Technology,
AlbaNova, SE-106 91 Stockholm, Sweden}
\date{\today}

\begin{abstract}
The realization of a directed current for a quantum particle in
a flashing asymmetric potential is studied. It is found that
a positive current, i.~e.\ in the direction
expected for a conventional diffusive ratchet, can be attained at
short times in the limit where the potential is weak and
quantum diffusion dominates,
while current reversal is obtained for stronger potentials.
A single parameter, the ratio between the kicking 
frequency and the optical lattice potential strength, determines 
both the degree to which quantum effects dominate, and the 
possibility of obtaining a ratchet current.
The effect should be readily observable in experiments.
\end{abstract}


\maketitle

The study of nanoscale quantum pumps is of considerable current
interest, motivated by
both the quest to explore basic quantum phenomena, by recent significant
experimental progress, and
also by the potential practical implications in terms of artificial devices
\cite{reimann2002a}.
Ratchet models have been introduced and studied
for processes that allow extraction of work from diffusive motion without
application of macroscopic gradients.
For example, ratchets have been discussed in the context of
biological motors \cite{astumian1997a} 
and for manipulating magnetic flux in artificially
nanostructured superconductors \cite{wambaugh1999a}.

With the advent of optical
lattices, the possibility has opened up to study dynamics of systems
of quantum particles in the absence of external noise.
These provide a novel setting for ratchet effects and interest
in purely quantum ratchets has therefore grown recently.
In rocking ratchets, where an alternating bias field is imposed on
the system, current reversal due to quantum effects has been
observed \cite{reimann1997a,linke1999a}. In an optical lattice,
transport was observed in a setup that utilizes oscillations
between internal states of the atoms to create an effective ratchet
potential that alternates between two configurations
\cite{mennerat1999a}.
Another realistic experimental setup for atoms in an optical
lattice is a flashing ratchet, i.e., an asymmetric potential
that is turned on and off periodically in time.
Reference \cite{monteiro2002a} carried out a theoretical study of
such a system in the case of
a small effective Planck constant, and found that no long-time
transport was
present in the case of strictly periodic
flashing. They devised instead a scheme with unequal spacing between
the flashes, which was shown to result in greatly enhanced transport.
Reference \cite{duffy2004a} studied a similar system
experimentally, but without any asymmetry in the potential,
and found at {\em short} times a greatly enhanced spreading in
momentum space due to
quantum resonances at certain parameter values (however, because of
the geometry, no directed transport could be attained).

In this paper we study a nonlinear quantum pump acting on a particle
trapped in an optical lattice,
driven by a flashing ratchet mechanism.
The focus of the paper is to determine the conditions for directed 
transport in a
one-dimensional flashing ratchet, in the regime where quantum
effects are appreciable. The aim is to sort out the dependencies on
the various parameters of the system and to provide physical
explanations for the parameter dependences and the absence of
long-time transport.

{\it System --}
Consider a particle in one dimension subject to the
Schr{\"o}dinger equation \cite{monteiro2002a,duffy2004a}
\beq\label{schrodinger}
i\heff\frac{\partial \psi}{\partial t} =
-\frac{\heff^2}{2}\frac{\partial^2\psi}{\partial x^2} +
K v(x) \sum_{n=0}^{\infty} \delta(t-n)\psi,
\enq
where
\beq
v(x) = \sin x + \alpha \sin 2x
\enq
is the ratchet potential, which we assume to be composed of two
standing laser waves. 
The parameter $\alpha$ controls the skewness of the periodic potential
that is the origin of the ratchet effect.
A value of $\alpha$ between 0 and 0.5 approximates a 
sawtooth potential where the sawteeth lean to the left;
this is expected to yield a positive current in the usual case
of classical diffusive motion. The profile is
more asymmetric for larger values of $\alpha$.
The potential is assumed to be flashed on and off at
periodic intervals. To facilitate
calculations we have taken the pulses to be delta functions in time.
The case $\alpha=0$,
corresponding to a pure sine potential, is formally equivalent to
the kicked rotor and has been extensively studied
theoretically and experimentally 
\cite{duffy2004a,daley2002a,izrailev1980a,book}.
The addition of a second sine wave adds complexity
to this classic problem and opens for the possibility of a ratchet
effect \cite{monteiro2002a}.

The units in Eq.\ (\ref{schrodinger}) are chosen such that the
temporal period of the
flashing and the spatial period of the lattice are unity. In
terms of physical quantities, the effective Planck constant is
\beq
\heff = 8\omega_R T,
\enq
where $T$ is the period of the flashing and $\omega_R=\hbar k_L^2/2m$
is the recoil frequency of the applied laser field, with $k_L$ the
recoil momentum 
giving a lattice
period $(2k_L)^{-1}$ for the optical potential felt by the atoms.
The effective potential strength is
\beq\label{kdef}
K = \heff \frac{T E_{\rm OL}}{\hbar} = \heff P,
\enq
where $E_{\rm OL}$ is the amplitude of the periodic potential of the
laser beam with the lower frequency.
We have implicitly defined the
auxiliary quantity $P = K/\heff$ in Eq.\ (\ref{kdef}), because as
we shall see, it determines much of the physics of the system.
In addition, $P$ has a simple interpretation: it is just the
ratio between the potential strength and the kicking frequency.

When the potential flashes are assumed to be delta-function kicks,
the time evolution can in principle be described analytically: in
momentum space the wave function at time $t$ is given by
\beq\label{timeevol}
\psi(k,t) = e^{-i\heff k^2/2}\sum_{q=-\infty}^{\infty}
F(k-q)\psi(q,t-1),
\enq
where the wavenumber difference $q-k$ is restricted to
integer values because of the periodicity of the potential, and
the function $F$ is given by \cite{monteiro2002a}
\beq\label{besselsum}
F(k) = \sum_{p=-\infty}^{\infty} J_{k-2p}(P)
J_p(\alpha P).
\enq
$J_n$ are Bessel functions. 
The dependency of $F$
on $P$ and $\alpha$ is suppressed for brevity. In the
kicked-rotor case, $\alpha=0$, all terms except $p=0$ vanish
in (\ref{besselsum}) and there results $F(k)=J_k(P)$.
Although the analytical formula,
Eqs.\ (\ref{timeevol}-\ref{besselsum}), is
useful for studying analytical properties,
it is easier to solve for the time evolution numerically using
a FFT algorithm. We assume an initially homogeneous wave function
with zero momentum. The time evolution will because of the
periodicity bring in only momentum components $\psi(k)$ 
with integer $k$.
The wave function is propagated in time, one period in each step,
by first multiplying the wave function components in $k$ space
by a phase factor $\exp(-i\heff k^2/2)$, then Fourier transforming
the wave function and applying the phase factor associated with
the potential, $\exp(-iPv(x))$, to the real-space wavefunction
before Fourier transforming back. 

\begin{figure}
\includegraphics[width=\columnwidth]{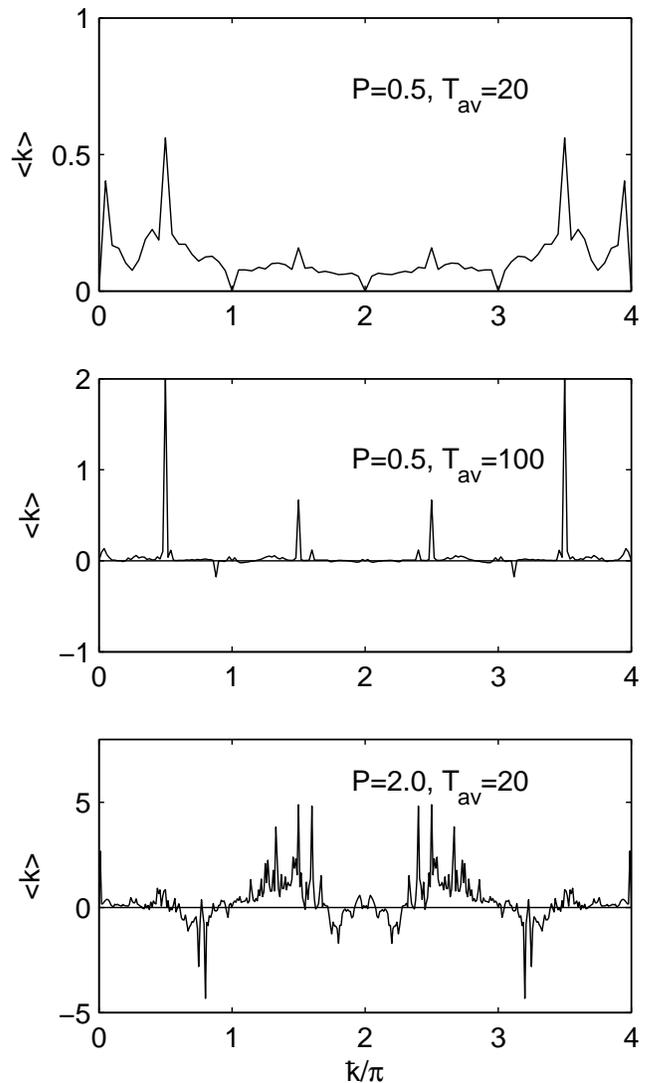}
\caption[]{Mean wave number of the wavepacket over the first
$\tav$ kicks, as a function of the effective Planck constant
$\heff$. The asymmetry of the ratchet potential has been chosen
to $\alpha=0.3$. Potential strength $P$ and averaging
time $\tav$ are as indicated in each panel.
\label{fig:overview}}
\end{figure}

{\it Effective Planck constant -- }
Figure \ref{fig:overview} displays the
mean wave number $\langle k \rangle$
of the wavepacket over the first $\tav$
temporal periods, as a function of the effective Planck constant
$\heff$.
The magnitude of the
$\sin 2x$-term in the potential is chosen as $\alpha=0.3$.
The main features of the wave number-vs-Planck constant curve do not
depend critically on the value of $\alpha$, but
as a rule the drift is stronger for larger $\alpha$.
We have held $P$ constant, so that
the potential strength in each of the panels of
Fig.\ \ref{fig:overview} varies along the abscissa as $K=P\heff$.
Consider first the topmost panel, where $P=0.5$ and $\tav=20$.
The curve is seen to be
dominated by a series of resonances, but overall there is positive
transport in the anticipated direction. Shown is only the curve for
$0 < \heff < 4\pi$, because it
is periodic in $\heff$ with period $4\pi$. When $P$ is kept
constant, the potential-energy part of the
time evolution operator does not depend on $\heff$, but only
the kinetic term $\exp(\heff \hat{k}^2/2)$ does,
and since spatial periodicity restricts the values of $k$ to
integers, the change $\heff \to \heff + 4\pi$ does not affect the
dynamics. It is thus not obvious that $\heff$
should be thought of as the parameter that determines whether the
system is classical or quantum mechanical.
Moreover,
according to Eq.\ (\ref{timeevol}), changing $\heff$ to $-\heff$ or
$4\pi-\heff$ only results in an overall complex conjugation of
the wave function and no change of the physics. This explains
the reflection symmetry of the curve about the point $\heff=2\pi$.

The resonances in $\heff$ for integer and half-integer multiples of 
$\pi$ are well known from the kicked-rotor case, $\alpha=0$ 
\cite{duffy2004a,daley2002a,izrailev1980a}. The algebra pursued 
in Refs.\ \cite{daley2002a,izrailev1980a} can be carried over to 
the present, asymmetric case and yields similar results. Details 
of the calculation will be presented elsewhere \cite{inpreparation}.

{\it Long-time behavior -- }
The second panel in Fig.\ \ref{fig:overview} shows the mean wave
number over an averaging time of $\tav=100$ kicks instead of 20.
Only the resonances remain in this
curve; away from these the mean wave number
has averaged out to nearly zero.
Apparently, no transport can be expected
at long times, except at resonances. This was also found in
Ref.\ \cite{monteiro2002a} and the reason is quantum
localization. In the kicked rotor problem, it is well known that
a quantum system will at short times undergo diffusion in
momentum space, just like in the classical case, at a diffusion
rate $D=K^2/2$.
However, after a so-called break time $t^* \propto D/\heff^2$, the
quantum time evolution ceases to follow the corresponding classical
evolution and the momentum distribution does not broaden any
more. This is analogous to Anderson localization, although the
phenomenon takes place in momentum space and not in real space
\cite{book}. The localization persists also in the case of an
asymmetric potential, $\alpha \neq 0$, and seems to forbid a persistent current
(see, however, Ref.\ \cite{monteiro2002a}). Observe that with
our definitions, the break time is simply $t^* \propto P^2$. In
other words, a large $P$ means a more pronounced quantum-classical
correspondence. The relevant parameter
determining the quantum/classical character of the dynamics is thus
$P$ and, perhaps somewhat counterintuitively, not $\heff$.

\begin{figure}
\includegraphics[width=\columnwidth]{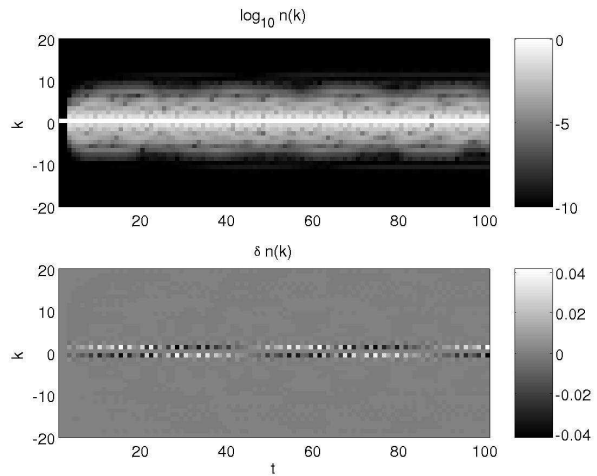}
\caption[]{Time evolution of the wave function in momentum space.
Upper panel displays the density distribution; the logarithm
is taken in order to enhance visibility. Each vertical
cross-section of the plot represents the squared wave function
$n(k,t)=|\psi(k,t)|^2$ at a time instant $t$.
The lower panel shows the asymmetry in the momentum
distribution, $\delta n(k,t) = n(k,t) - (n(k,t)+n(-k,t))/2$.
\label{fig:kwavefunction}}
\end{figure}

The present parameter value, $P=0.5$, is far into the quantum
regime since it results in a break time $t^*$ less than one, and 
thus no quantum-classical correspondence can be expected. 
Figure \ref{fig:kwavefunction} displays the time evolution of the
wave packet for a representative choice of
parameters away from
resonances: $\heff=1.7$, and as before, $P=0.5$ and $\alpha=0.3$.
The lower panel of Fig.\ \ref{fig:kwavefunction} shows the
momentum asymmetry, defined as the deviation of the density
from its symmetrized value:
$\delta n(k) = n(k) - (n(k)+n(-k))/2$. It is seen that the density
distribution initially broadens, with a slight tendency towards
positive $k$, whereafter the width of the distribution saturates
due to localization, because of the short break time. 
From that point, the momentum distribution
begins to oscillate back and forth within the wave packet, yielding
a zero net momentum at longer times. The initial tendency towards
positive $k$ is in accordance with our expectancies for a ratchet;
this is further discussed below.

\begin{figure}
\includegraphics[width=\columnwidth]{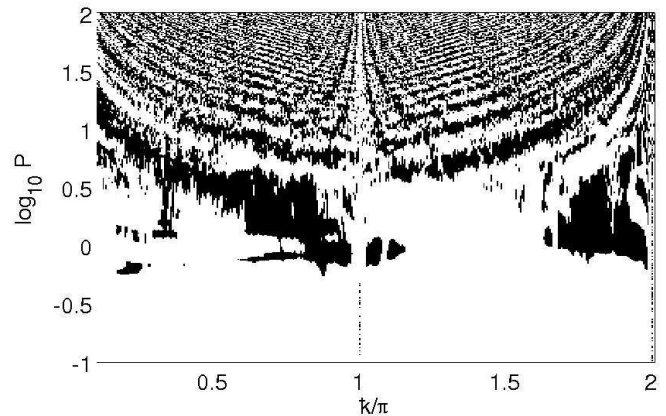}
\caption[]{Regimes of positive (white areas) and negative
transport (black areas) for the quantum ratchet with
$\alpha=0.3$. The mean momentum is taken over the first
$\tav=20$
periods. The first and third panels in Fig.\ \ref{fig:overview}
correspond to the horizontal cross-sections of the present curve
at constant $P=0.5$ and $P=2$, respectively, while the middle
panel of Fig.\ \ref{fig:overview} is taken at a different
$\tav$.
\label{fig:negative}}
\end{figure}

{\it Kick strength -- }
As discussed above, it is customary to discuss the physics of
the present system in terms of the kick strength
$K$, because the Schr{\"o}dinger equation
(\ref{schrodinger}) assumes the intuitively expected form, 
and moreover, the effective Planck constant $\heff$ assumes the 
role of a quantum parameter.
However, as we have seen, $P=K/\heff$ is in fact the natural
parameter that determines the degree to which the time evolution
is quantum mechanical, and furthermore, as we now show,
it turns out to determine whether a short-time ratchet effect is present.
In the third panel of Fig.\ \ref{fig:overview} we report
the result for slightly stronger kicks than in the examples above, 
$P=2$. The average is
taken over twenty periods.
For this kick strength, the curve has
a considerably more broken appearance. In particular, the drift
is no longer positive for all values of $\heff$: it is just
as likely to be negative. Indeed, we do find that for small
values $P \lesssim 1$, the drift is always positive, except at
anti-resonances, but it can be negative for larger $P$.
Figure \ref{fig:negative} shows the regions in $P$-$\heff$ space
where the drift over the first twenty kicks is positive and
negative, respectively.
It is seen that for
large values of $P$ or $K$ and for $0<\heff<\pi$, the black
regions run along lines of constant $K$, indicating that in
the classical regime, the dynamics depends on $K$ rather than
$P$. On the other hand, what is interesting here is the dividing 
line below which the drift is purely positive, and that is 
clearly a line of constant $P$, not $K$.

The observation of negative transport (current reversal) for
quantum ratchets in the zero-temperature limit has in the case of
a {\em rocked} ratchet been explained with a tunneling argument
\cite{reimann1997a}. That argument is not applicable for a
delta-function {\it flashing}
ratchet; obviously another mechanism is at play here.
Consider the time evolution of a wavepacket
originally concentrated within a well. Between two
kicks, it undergoes diffusion according to the Schr{\"o}dinger
equation (\ref{schrodinger}) and diffuses out to a final
position uncertainty $\Delta x_{\rm diff} \sim \heff$ in the
present units.
The particle also
receives an impulse from the periodic potential, which is
of the order $\Delta p \sim K$ if the particle initially resided on
one of the slopes of the potential, and hence travels a distance
comparable to $\Delta x_{\rm ballistic}\sim K$ in one period.
We now see that the condition
for a ratchet effect at short times,
\beq
P \lesssim 1,
\enq
translates into a condition on the physical time scales,
\beq
\Delta x_{\rm ballistic} \lesssim \Delta x_{\rm diff}.
\enq

In summary, we have investigated the conditions for
directed transport for a quantum particle in a flashing
ratchet potential.
It is found that at short times, a directed current will flow
in the anticipated direction if the kick strength $P$, defined
as the ratio between the amplitude of the potential and the
flashing frequency, is less than unity; this coincides with
the limit in which the motion is inherently quantum mechanical,
and quantum diffusion dominates over ballistic transport.
It would be interesting to look for these findings in 
experiment:
the levels of precision, 
noise and parameter control attainable in current experiments 
\cite{mennerat1999a,duffy2004a} are certainly sufficient 
to test the predictions made in the present paper.


\end{document}